# Transfer Learning Based Diagnosis and Analysis of Lung Sound Aberrations

Hafsa Gulzar[1], Jiyun Li[1]*, Arslan Manzoor[2], Sadaf Rehmat[3], Usman Amjad[4] and Hadiqa Jalil Khan[4]

[1]Computer Science and Technology, Donghua University, Shanghai, China
hafsa@mail.dhu.edu.cn  jyli@dhu.edu.cn
[2]Mathematics and Computer Science, University of Catania, Catania, Italy
arslan@mail.dhu.edu.cn
[3]Computer Science Department, PIEAS, Islamabad, Pakistan
Sadafrehmat123@gmail.com
[4]Computer Science and Information Technology, Islamia University of Bahawalpur, Pakistan
muhammadusmanamjad1@gmail.com  hadiqajalil10@gmail.com

*ABSTRACT*

*With the development of computer -systems that can collect and analyze enormous volumes of data, the medical profession is establishing several non-invasive tools. This work attempts to develop a non-invasive technique for identifying respiratory sounds acquired by a stethoscope and voice recording software via machine learning techniques. This study suggests a trained and proven CNN-based approach for categorizing respiratory sounds. A visual representation of each audio sample is constructed, allowing resource identification for classification using methods like those used to effectively describe visuals. We used a technique called Mel Frequency Cepstral Coefficients (MFCCs). Here, features are retrieved and categorized via VGG16 (transfer learning) and prediction is accomplished using 5-fold cross-validation. Employing various data splitting techniques, Respiratory Sound Database obtained cutting-edge results, including accuracy of 95%, precision of 88%, recall score of 86%, and F1 score of 81%. The ICBHI dataset is used to train and test the model.*

## 1. INTRODUCTION

The sounds of the lungs are vital indications of respiratory health and disease. In recent years, lung sound diagnosis and classification have attracted much research attention [1-7]. Breathing is so necessary that in 24 hours, an average human can breathe 25,000 times [8-12]. According to the World Health Organization (WHO), the COVID-19 epidemic on May 24, 2021, there have been 166,860,081 verified cases, with 3,459,996 deaths recorded [13]. Heart disease is a significant issue; 610,000 people died because of heart disorders, and lung disease is the leading cause of heart problems. Lung disorders are becoming the world's second leading cause of death. The acute and chronic respiratory conditions that directly impact individuals' health include COPD, asthma, lung inflammation, infectious diseases, tuberculosis, pulmonary embolism, sleep apnea, and occupational lung diseases [2, 14-26].

Effective respiratory disease management requires early diagnosis and patient monitoring. It is easy to discern lung sounds mainly through deep learning technologies. This system is suitable for removing general noise effects and classifying lung sounds into two categories:

usual and unexpected. The most common unexpected lung sounds noted above normal ones are crackles, wheezes, and squawks, and their presence usually implies a pulmonary problem [27]. Speech recognition was boosted via neural network model (acoustic signal source identification). A problematic human-level categorization performance was demonstrated.

This approach employed MFCC as a pre-processing module. The MFCC method is used to convert signals into spectral images. ResNet101 and VGG16 are two feature extractors with DCNN classifiers [28-30]. This approach incorporated CNN classification, respiratory sound, and pre-trained image recognition. Additionally, we contrasted the outcomes of these feature extractors with RNN models such as the LSTM and BLSTM (Bidirectional Long Short-Term Memory). The NMADCNN end-to-end hybrid DCNN strategy categories lung sound as normal, wheezes, crackles or both and identifies noise in breath cycles [7, 9, 31, 32]. So, in this study, DCNN classifier ResNet101 and VGG16 are improved, which outperformed all competing algorithms compared to LSTM, BLSTM, and many other models. The dataset ICBHI 2017 Respiratory Sound Database includes normal and three other types of inadvertent lung sounds, such as wheezes, crackles and their combination. These sounds are used to train and evaluate the model [33].

**Significance:** Numerous medical and computer science disciplines might benefit from this study. This work helps determine the appropriate CNN model architecture for lung sound detection and classification and compares it with other CNN and RNN models.

**Objectives:** In this article, normal sound, wheeze, crackles, and wheeze with crackles are proposed as four categories of lung sounds.

The following are the primary goals of our research:
- To transform audio impulses into a visual layout using the MFCCs approach.
- CNN models, as well as CNN's updated models and methods like VGG16 and Res101, are used to specify implementation and results.
- Try to get more than 95% accuracy by improving our models.
- The results are compared by RNN models such as the LSTM and BLSTM to determine the optimal architecture for lung sound detection and classification.

**Statement of problem**: The real problem is to improve the Lung sound aberration results by using CNN models (ResNet101 and VGG16) and compare it with different models.

**Structure of the paper:** Section 2 and 3 presents a relevant literature history and methodology including dataset, pre-processing and modeling. Section 4 and 5 contains our experimental approach and outcomes. Finally, section 6 elaborates the conclusion.

## 2. RELATED WORK HISTORY

The precise identification of lung disorders such as Covid-19 and viral/bacterial pneumonia has recently gained considerable interest. One of the studies in this field used statistical signal processing approaches to classify breath sounds into three distinct categories: namely soft, soft, and loud [7]. Nevertheless, it does not imply the possibility of using parameters other than moderate, frequent, and severe breath counts for categorization purposes. In another significant study, convolutional neural networks are used for the automated classification of heart and lung sounds. This study provides a solid framework for the application of Machine Learning and Artificial Intelligence to lung auscultations [5]. This research aimed to combine respiratory sound, CNN classification, and time-series feature extraction from pre-trained images. We also assessed the performance of several feature extractors. The LSTM model is utilized to obtain the findings after the MFCC model has been used for feature extraction [12]. The fact that LSTM outperforms a few competing models with an ICBHI score of 74% after being pitted against them shows the power of the LSTM-based framework in lung sound data pre-processing. This research aimed to combine respiratory sound, CNN classification, and time-series feature extraction from pre-trained images. We also assessed the performance of several feature extractors [34]. The LSTM model is utilized to obtain the findings after the MFCC model has been used for feature extraction [35]. The fact that LSTM outperforms a few competing models with an ICBHI score of 74% after being pitted against them shows the power of the LSTM-based framework in lung sound data pre-processing [2, 30].

## 3. METHODOLOGY

In addition to outlining the measures used to assess the effectiveness of the built neural network, we also present the methodology utilized in this article and section. The method's significant components are depicted in the diagram in Figure 1. Data pre-processing, feature extraction using VGG16 (transfer learning), classification using CNN with training and testing, prediction using 5-fold cross-validation, and performance analysis utilizing metrics relevant to the current position make up its four main modules.

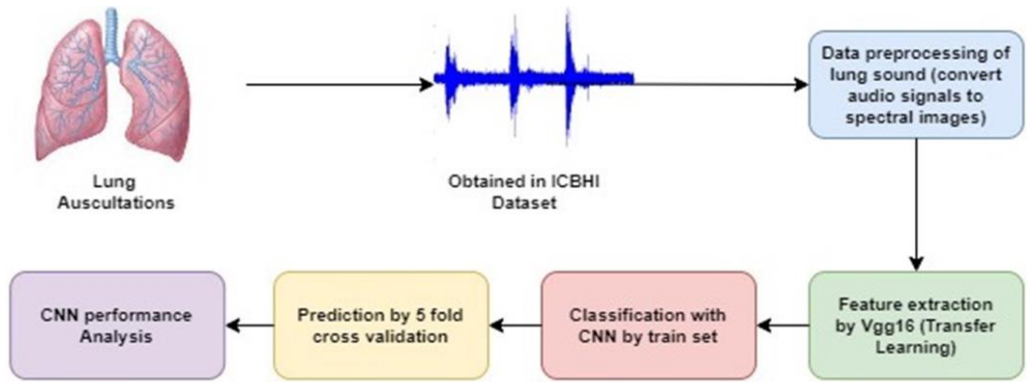

**Fig. 1. Description of our models.**

### 3.1. Data Set

Rocha et al. data set from the Challenge ICBHI 2017 was used in this article [27]. The primary purpose of this collection of breathing sounds was to aid the Informatics in Biomedical Health ICBHI 2017 research challenge. The database, created by two research teams in Portugal and Greece, contains 920 recordings from 126 subjects that total 5.5 hours in length. 6,898 respiratory cycles were recorded, of which 3,642 were abnormality-free, 1,864 had crackles, 886 had wheezing, and 506 had both. The recordings, which were recorded using a variety of equipment, varied in length from 10 to 90 seconds. In addition, the places at which the recordings could be obtained were detailed in the article. Data is collected from seven sites, recorded both chest sound and breathing sounds. These recordings were collected from the real patients. Patients with bronchiectasis, asthma, COPD, respiratory infections, and lower respiratory diseases were included. Both clinical and non-clinical settings (patients' homes) had their sound recordings made. Patients range in age from young children to older adults.

Three skilled medical professionals, including two pulmonologists with specialized training and one cardiologist, documented the respiratory sound characteristics in the database.

**Table 1: Total number of dataset cycles for ICBHI 2017.**

| Dataset | Total |
|---|---|
| Crackles cycles | 1864 |
| Wheezes cycles | 886 |
| Combination of crackles and wheezes | 506 |
| Normal cycles | 3642 |
| Total | 6898 |

### 3.2. Pre-processing

During processing, the 5-second-long audio recordings are windowed or separated into parts. If required, segments are filled with zero to ensure that their sizes are same. The number of samples from each class can be used for CNN training and can be increased using this technique. As a result, there are the following numbers for each category: There are 7,488 samples for wheezing, 6,415 samples for crackles, 732 samples total for both classes (crackles and wheezing), and 6,850 samples total for the class without any respiratory abnormalities.

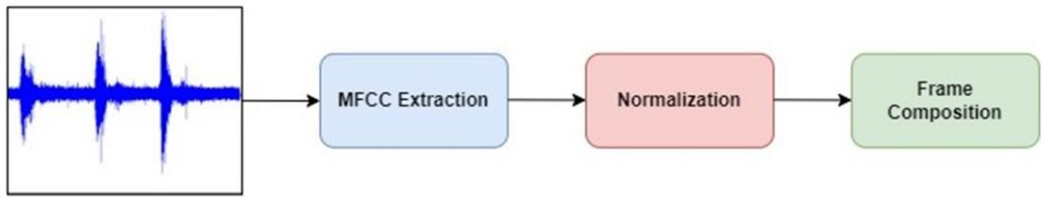

**Fig. 2. All steps of pre-processing modules.**

Spectrograms display the frequency spectrum of a sound and how it varies over a short time. The primary distinction is that an MFCC utilizes a quasi-logarithmic spaced frequency scale, which is more similar to how the human auditory system perceives sound, as opposed to a spectrogram, which uses a spaced linear frequency scale (such that an equal number of Hertz are used to space out each frequency section) [36, 37].

**Table 2: Cycle data for an ICBHI 2017 database audio file.**

| Cycles | Start time | End time | Crackles | Wheezes |
|---|---|---|---|---|
| 1 | 0.804 | 3.256 | 0 | 0 |
| 2 | 3.256 | 5.566 | 0 | 0 |
| 3 | 5.566 | 7.851 | 0 | 1 |
| 4 | 7.851 | 10.054 | 0 | 1 |
| 5 | 10.054 | 12.066 | 1 | 0 |
| 6 | 12.066 | 14.47 | 1 | 0 |
| 7 | 14.47 | 16.696 | 1 | 1 |
| 8 | 16.696 | 18.887 | 1 | 1 |
| 9 | 18.887 | 19.792 | 1 | 1 |

### 3.3. Modeling

Author Mel-spectrograms of lung sounds are generated, and VGG16 is used to extract features. The classification is made using CNN, while the prediction is performed using 5-fold cross-validation. Two or three breathing cycles are needed to accurately analyze lung sounds. The typical normal respiratory rate is 15 to 20 breaths per minute (three to four seconds each breath), while in pathologic circumstances, it tends to be faster. So, after experimenting with several choices, six seconds ultimately chosen for the duration of the breathing sound. In the process of transfer learning, pre-trained models such as VGG16 are used as feature extractors. The 16-layer VGG16 model, which is trained on fixed-size pictures and uses small-size kernels with a responsive Feld 33, processes the input via a series of convolutional layers. Our model's input size is 256 x 256, while VGG16's default input size is 224 x 224. (Fig. 3). Without introducing a fully linked layer and by freezing all five convolutional blocks, we can use weights that had already been trained on ImageNet to predict the test sets using a single layer of a simple CNN. The 5-fold cross-validation used method to prevent overfitting (Fig. 4). The dataset is randomly divided into training sets of 80% and test sets of 20%, with 20% of the training set used for validation.

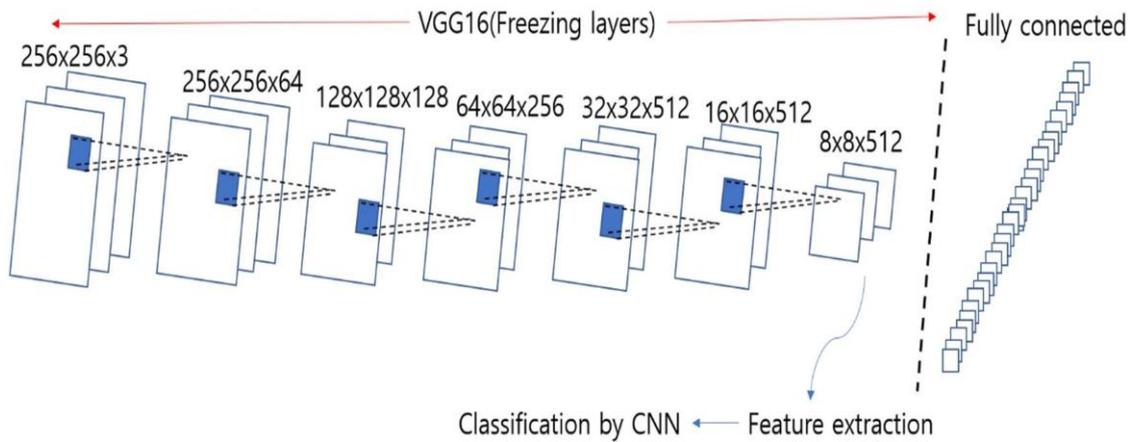

**Fig. 3. Feature extracted by VGG16 architecture.**

The VGG16 architecture is used in our model. Our model's input size is 256×256. Without a completely linked layer, all layers are constrained to extracting characteristics and classifying respiratory sounds.

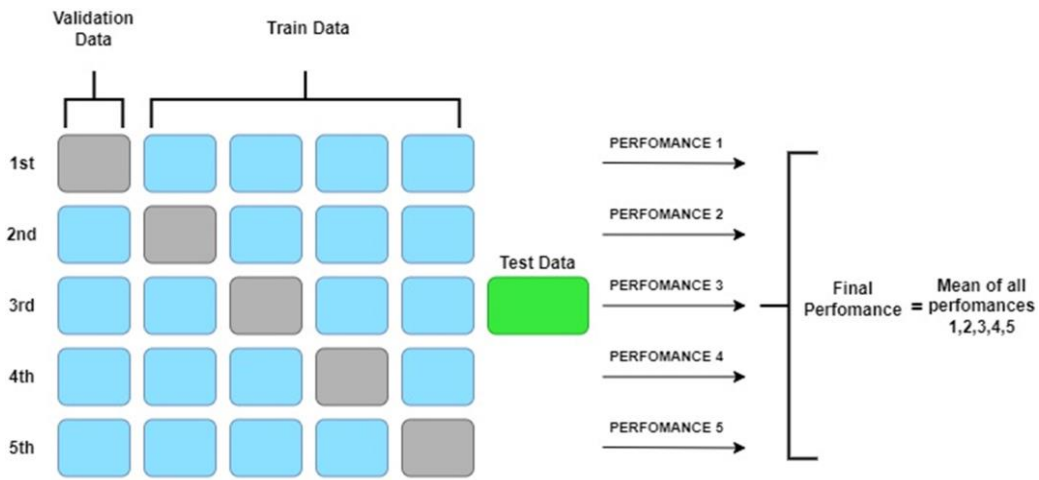

**Fig. 4. Cross-validation with five levels of replication. The outcomes of five iterations are the average of all performances.**

The main premise of 5-fold cross-validation is to split the training set into 5 sections. The model is trained to use four of the five divisions each time, and one is used to assess it. As a result, the data set's instances are used once for testing and four times for training. Averaging the measurements determines the result. From the results of our models, accuracy, precision, recall score extrapolated.

For deep feature extraction, we employ the completely connected layers, each of which produces an output in 4096 dimensions. The class labels in number 10 are all predicted using the SVM method. Transfer learning is applied to the pre-trained VGG16 model in Fig. 2. The input lung spectrogram images were employed to further trained (i.e., fine-tuned) the pre-trained VGG16 model. The VGG16 model's final three layers are not considered to achieve fine-tuning flexibility since the layers are set up for 1,000 classes of ImageNet tasks.

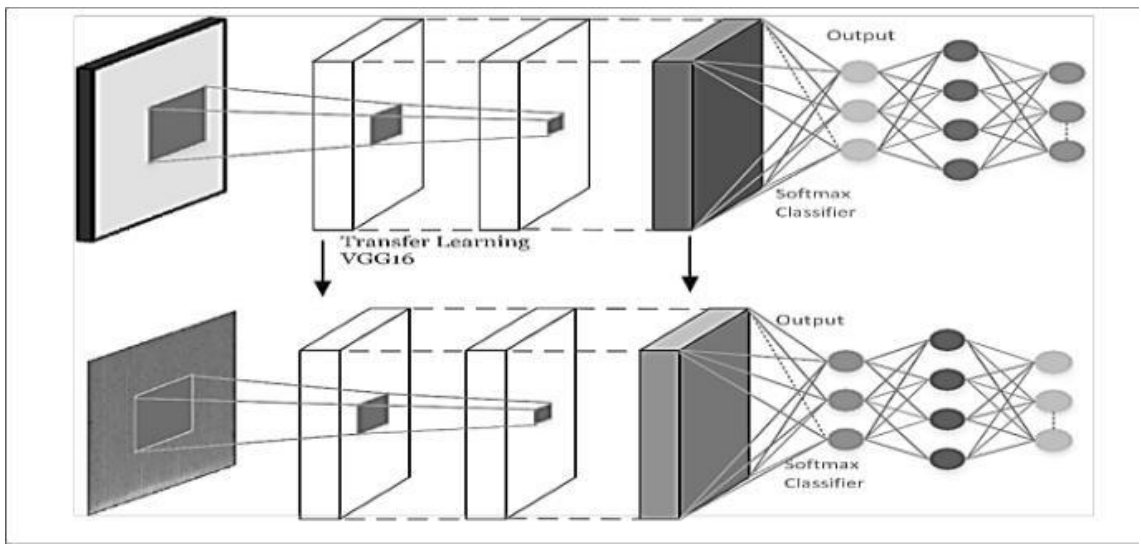

**Fig. 5. The suggested transfer learning approach for classifying lung sounds.**

## 4. EXPERIMENT

The study started with LSTM, BLSTM, random forest, focal function, and SVM-based classifiers to compare them to VGG16 and ResNet101 and the CNN model.

### 4.1. Extra Features

The extraction of the resources needed to train our model is covered in the next stage. To do this, high-precision image classification techniques have been used to discover classification features using a visual representation of each audio sample. MFCCs make it more proficient. Each audio sample has a visual representation according to the resources we extracted using the MFCC for each audio file in the dataset. In this manner, these images used to train the classifier.

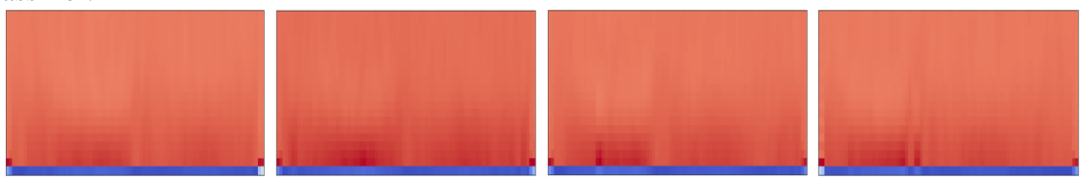

**Fig. 6. MFCC Normal Class.**

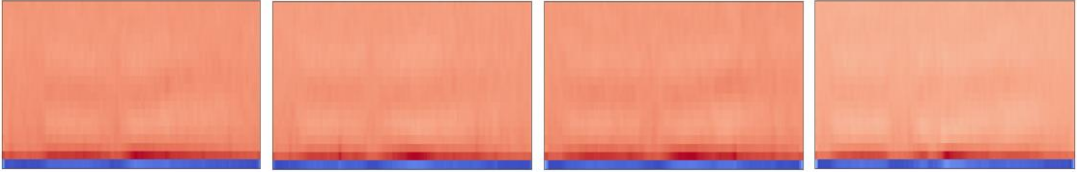

**Fig. 7. MFCC Wheezes Class.**

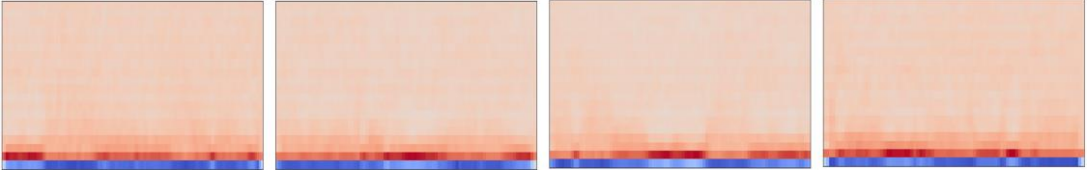

**Fig. 8. MFCC Crackles Class.**

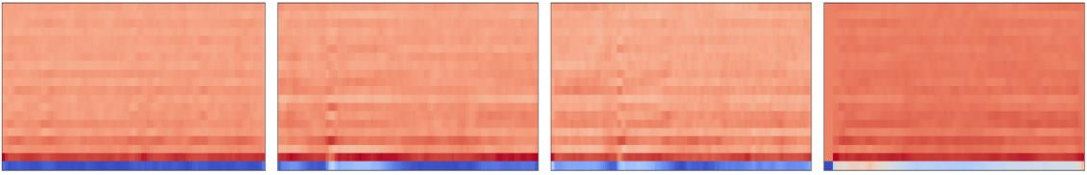

**Fig. 9. MFCC Both Class.**

### 4.2. Evaluation and Assessment Criteria

The statistical indications of a diagnostic test's ability to identify aberrant from regular patients are unreliable because of the dataset, sensitivity, and specificity. The original paper's data set includes suggestions for a sensitivity, specificity, and overall score [38-41]. The general assessment result is as follows:

$$\text{Score} = \frac{\text{Sensitivity} + \text{Specificity}}{2} \quad (1)$$

Cross-validating patients allowed for the evaluation of results. Cross-validation entails splitting the dataset into separate iterations for training and testing. The cross-validation result is reliable as a result. There are no trained group patients in each split test set since 5-fold cross-validation is being used in this case. This was verified using actual, real-oriented data partitioning.

### 4.3. Experimental Setup

There have been many experiments that have used different data and pre-processing phases. Comparing different ways to evaluate their accuracy and effectiveness is the basic idea. Respiratory noises from patients with pulmonary illnesses are rectified. Pulmonologists verified and categorized the sounds. Mel-spectrogram created from the sounds, and VGG16 is used to extract the features (transfer learning). According to CNN, respiratory sounds are categorized. Deep learning classification of respiratory sounds could be useful for pulmonary illness screening, monitoring, and diagnosis.

Although it is an excellent and robust experiment, it lacks an endwise experiment since each lung sound must be divided into the lung cycle before the lung cycle technique can be applied to it. The third experiment is directed due to the second experiment's flaw. Here, model effectiveness is tested to locate important data and where to place it inside the vast feature space. It necessitates scoring the lung-based data for dualistic components.

There are formulas for precision, false alarm rate, sensitivity, and specificity, respectively, in equations 2 to 5.

$$Sensitivity = \frac{True\ Positive}{True\ Positive + False\ Negative} \quad (2)$$

$$Specificity = \frac{True\ Negative}{True\ Negative + False\ Positive} \quad (3)$$

$$SpecifFalse\ Alarm\ Rate = 1 - specificity \quad (4)$$

$$Precision = \frac{True\ Positive}{True\ Positive + False\ Negative} \quad (5)$$

Accuracy is defined as the proportion of properly recognized participants to the full set of subjects.

## 5. RESULTS

Implementation for the modeling is done in the TensorFlow framework using multiple supporting python libraries for image processing. Our training is done on NVIDIA titan GPU with an Intel 9[th] generation i7 processor with 64GB memory. The model is trained in batches. In the system, raw sound is taken as input files and separated them into four separate classes. Extracted MFCCs from these sound files for training our model. The output layers of the VGG16 model and configure other dimensions according to our problem. After training, to test our model used the testing dataset.

Hyperparameter tuning is the key step in improving the extracted features and fine-tuning the model according to the specific problem. Results have been enhanced in hyper-parameter tuning during the 5-fold cross-validation. The results are significantly better due to the model's fine-tuning and MFCCs' noise elimination.

**Table 3: The metrics used to assess how effectively the neural network test performed.**

| Class | Accuracy | Precision | Recall | F1 Score |
|---|---|---|---|---|
| Normal | 97.46% | 0.98 | 0.97 | 0.98 |
| Wheezes | 97.68% | 0.90 | 0.92 | 0.91 |
| Crackles | 96.96% | 0.94 | 0.95 | 0.94 |
| Both | 97.9% | 0.86 | 0.85 | 0.86 |

**Table 4: Confusion matrix**

|  | Normal | Wheezes | Crackle | Both |
|---|---|---|---|---|
| Normal | 706 | 05 | 5 | 4 |
| Wheezes | 05 | 163 | 9 | 4 |
| Crackle | 10 | 6 | 356 | 7 |
| Both | 06 | 03 | 5 | 86 |

In our model, all the classes perform almost equally well. Using transfer learning and Softmax activation function in the output layer after pre-processing from the dataset outperformed all other models with an overall accuracy of 95.06%.

**Table 5: The classification accuracy for the CNN models ResNet-50 and AlexNet**

| CNN models | Transfer learning +SoftMax (Acc%) |
|---|---|
| AlexNet | 60.5 |
| ResNet-50 | 59.10 |
| **VGG16(our)** | **95.06** |

VGG16 received the highest score in Table 6 of 95% for each respiratory cycle. Following the completion of the test set calculation, a comparison is made between complex models and our suggested approach. The sensitivity and F-score of the method are shown in Table 6.

**Table 6: Individually breathing cycle Results**

| Feature Extractor | Accuracy | Precision | Recall score | F1 Score |
|---|---|---|---|---|
| InceptionV3 | 0.74 | 0.71 | 0.72 | 0.72 |
| ResNet101 | 0.78 | 0.74 | 0.73 | 0.73 |
| **VGG16** | **0.95** | **0.88** | **0.86** | **0.81** |

Table 7 demonstrates that by utilizing the aforementioned metrics. Suggested models outperform all their rival models in every category; the VG16 technique provided the best presentation of above 95%, 88%, and 86%. As evidenced by the 1-2% variation in the ICBHI score across various frame-making settings, the context based on CNN to the critical stage in the pre-processing of lung-sound information is potent or forceful. Since baseline models are not employed to balance data-damaging components, such as the sound of repetitive breathing, another model must deal with the large volumes and localization of the breathing cycle. Therefore, neither PCA nor augmentation contributes to addressing these issues.

**Table 7: Regarding F-score, accuracy, specificity, and sensitivity, compared the techniques.**

| Feature extractor | Accuracy | Precision | Recall score | F1 Score |
|---|---|---|---|---|
| InceptionV3 [20] | 0.74 | 0.71 | 0.72 | 0.72 |
| DenseNet201 [20] | 0.78 | 0.73 | 0.96 | 0.74 |
| LSTM-DAE [19] | 0.92 | 0.71 | 0.78 | 0.94 |
| BLSTM-DAE [19] | 0.93 | 0.74 | 0.80 | 0.97 |
| VGG19 [20] | 0.80 | 0.74 | 0.82 | 0.76 |
| LSTM [19] | 0.88 | 0.86 | 0.62 | 0.72 |
| BLSTM [19] | 0.90 | 0.89 | 0.73 | 0.74 |
| Boosted Tree [20] | 0.72 | 0.78 | 0.21 | 0.49 |
| ResNet101 | 0.78 | 0.74 | 0.73 | 0.73 |
| **VGG16 (Our)** | **0.95** | **0.88** | **0.86** | **0.81** |

## 6. CONCLUSION

One of the most significant concerns in public health, the automatic detection of lung disorders, is the focus of this effort. No complicated datasets including sounds, background noises, and a range of sampling frequencies have been employed for lung sound classification, despite the fact that there has been many research on the subject. The majority of the work was completed utilizing conventional techniques. Deep learning, a cutting-edge technique, is applied to the problem of lung sound detection to improve classification performance. Images with one-to-one spectrogram qualities are obtained with colormap during the pre-processing stage of the suggested approaches to extract deep features and apply them to finetune. The CNN VGG16 model is utilized to accomplish feature extraction in both deep learning techniques. Additionally, the CNN Alex Net and ResNet-50 classification models are evaluated, and the

VGG16 model is taken for the recommended methods since it provided superior classification accuracy.

Compared to ResNet101, which received a score of 0.78, and many other models, VGG16 achieved a score of 0.95, outperforming all competing approaches. However, it must be upgraded to produce even better outcomes.

## ACKNOWLEDGEMENTS

The author thanks' Prof. Jiyun Li for her continuous support during this research. Sincerely appreciate her for helping us at every step.

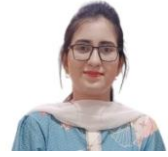

**Hafsa Gulzar**

Hafsa Gulzar received her Bachelor's degree in Computer Science from Islamia University of Bahawalpur, Pakistan. She is currently a Masters Student in Computer Science at the College of Computer Science and Technology, Donghua University, Shanghai, China. Her current research interests include Machine Learning, Data Mining and Artificial Intelligence.

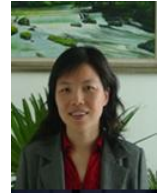

**Li Jiyun**

Li Jiyun is currently working as Professor at College of Computer Science and Technology, Director of Donghua Hesheng Joint Laboratory, Member of China Artificial Intelligence and Artificial Psychology committee and Member of China Graphics and Image Society. My current research interest are Data Engineering Machine learning, Psychological modelling of Artificial Intelligence, Decision Making, platform and Application Research and Development based on AI model.